\documentclass[a4]{article}

\usepackage{cite}

\usepackage{graphicx}
\usepackage{hyperref}

	 \usepackage{subfig} 
	 \usepackage{gensymb} 
	 \usepackage{color} 

\newcommand{\vJ}{{\bf J}}
\newcommand{\vE}{{\bf E}}

\newcommand{\vA}{{\bf A}}
\newcommand{\vH}{{\bf H}}
\newcommand{\vT}{{\bf T}}

\begin{document}

\title{\flushleft \bf Demagnetization of cubic Gd-Ba-Cu-O bulk superconductor by crossed-fields: measurements and 3D modelling\footnote{This article has been finally published with reference ``IEEE Trans. Appl. Supercond., vol. 28, iss. 4, 2018". The published version can be found at \url{https://doi.org/10.1109/TASC.2018.2808401}}}

\author{Milan Kapolka, Jan Srpcic, Difan Zhou, Mark D. Ainslie,\\Enric Pardo and Anthony R. Dennis\\
\normalsize{$^1$Institute of Electrical Engineering, Slovak Academy of Sciences,}\\
\normalsize{Dubravska 9, 84104 Bratislava, Slovakia.}\\
\normalsize{$^2$Department of Engineering, University of Cambridge,}\\
\normalsize{Trumpington Street, Cambridge, CB2 1PZ.}}

\maketitle

\begin{abstract}
Superconducting bulks, acting as high-field permanent magnets, are promising for many applications. An important effect in bulk permanent magnets is crossed-field demagnetization, which can reduce the magnetic field in superconductors due to relatively small transverse fields. Crossed-field demagnetization has not been studied in sample shapes such as rectangular prisms or cubes. This contribution presents a study based on both 3D numerical modelling and experiments. We study a cubic Gd-Ba-Cu-O bulk superconductor sample of size 6 mm magnetized by field cooling in an external field of around 1.3 T, which is later submitted to crossed-field magnetic fields of up to 164 mT. Modelling results agree with experiments, except at transverse fields 50\% or above of the initial trapped field. The current paths present a strong 3D nature. For instance, at the mid-plane perpendicular to the initial magnetizing field, the current density in this direction changes smoothly from the critical magnitude, ${J_c}$, at the lateral sides to zero at a certain penetration depth. This indicates a rotation of the current density with magnitude ${J_c}$, and hence force free effects like flux cutting are expected to play a significant role.
\end{abstract}

Keywords - hight temperature superconductors, numerical simulation, Finite element analysis, flux pinning, demagnetization, superconducting magnets
 

\section{Introduction}

Superconducting bulks are promising for permanent magnets. Bulks can trap higher magnetic field compared to ferromagnetic permanent magnets. However, they need to be cooled down below a certain critical temperature, ${T_c}$. The world record of the trapped field is 17.6 T at 26 K \cite{Durrell14SST}. There has been an important effort to develop such superconducting bulks \cite{Namburi16JECS}, \cite{Namburi16SST}. An important problem of superconducting bulks for many applications, such as motors, is demagnetization caused by longitudinal or transverse 
applied magnetic fields. Therefore, full 3D models are necessary which can reveal all demagnetization properties or finite size effects, while 2D cross-sectional models cannot. Superconducting cubic bulks 
present a higher ratio of superconducting mass to the free space, compared to other systems, such as stacks or cables of thin tapes. However, cubic samples are not solved, yet. 

There are many 3D modelling methods for superconductors, such as Finite Element Methods, FEM, and variational methods. There are several formulations of FEM like the ${\vH}$-formulation
\cite{Escamez16IES}, \cite{zermeno14SSTa}, \cite{grilli13Cry}, ${\vA-\phi}$ vector and scalar potential \cite{Farinon14SST}, ${\vT-\Omega}$ current and magnetic formulation \cite{Grilli05IES} or 
${\vH}$ formulation with homology-cohomology \cite{Stenvall14SST}. A completely different approach represents the variational methods. 
These also exist in several formulations in 3D like those for ${\vH}$ \cite{Bossavit94IEG,badia01PRL}, mixed ${\vH}$ and magnetic scalar potential $\psi$ \cite{Elliot06JMA}, 
and ${\vT}$ effective magnetization \cite{Pardo17JCP} (the latter also known as Minimum Electro-Magnetic Entropy Production in 3D, MEMEP 3D). Another practical formulation for 2D 
is the ${\vJ}$ formulation \cite{prigozhin97IES,Pardo15SST}. All FEM formulations and most varitional method formulations require to solve the surrounding air around the sample, in addition 
to the sample itself. Of all methods above, only the variational methods in the ${\vT}$ and ${\vJ}$ formulations avoid spending degrees of freedom in the air.

There are several studies of cross-field demagnetization. Those involve 2D FEM modelling using the ${\vH}$ formulation and experiments \cite{Vanderbemden03IES},\cite{vanderbemden07SST}, 
the study of hybrid (ferromagnetic/superconducting) structures by 2D and 3D modelling and experiments \cite{Fagnard16SST}, and the comparison of numerical calculation based on 
${\vA}$ and ${\vH}$ formulation (FEM) with theory of Brandt and Mikitik (thin strip) \cite{Campbell17SST}. 

Usually grown samples are cylindrical pellets. This motivated that all published works on 3D modelling and most experimental works are for this shape. However, many applications require to cut the 
original pellets into other shapes, such as rectangular prisms or cubes. At present, demagnetization by cross-field of a cubic sample is not well known. In this article, we study demagnetization 
of cubic bulks by experiment and by 3D modelling based on MEMEP 3D and ${\vH}$ formulation FEM. 


\section{Methodology of measurements}

\begin{figure}[tbp]
\centering
{\includegraphics[trim=0 0 0 0,clip,height=4.0 cm]{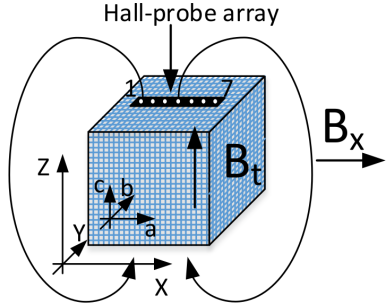}} 
\caption{Sketch of Hall probe sensors and direction of the applied fields, trapped field and ripples position relative to the sample.}
\label{Demag.fig}
\end{figure}

In this study we measured demagnetization of superconducting cubic bulks due to applied magnetic fields transverse to the trapped field. A 10wt\% Ag-containing GdBa$_2$Cu$_3$O$_{7-\delta}$ 
superconducting pellet was fabricated using the top-seeded melt growth (TSMG) process \cite{Namburi16JECS}. The cubic sample was cut down from a pellet with dimensions ${6.08\times6.04\times5.98}$ mm. 
We used the field-cool (FC) method, in order to magnetize the sample, which requires lower applied magnetic field than zero field-cool (ZFC) \cite{Ainslie15SST}. 
The FC method consists of the following steps:
\begin{itemize}
 \item The sample at room temperature, is inserted to the split coil electromagnet, which is ramped up to 1.3 T over 10 s.
 \item The sample is cooled down by liquid nitrogen over 15 minutes.
 \item The magnetic field is ramped down in the electromagnet over 100 s with rate 13 mT/s. The sample is magnetized parallel to the c-axis.
\end{itemize}      

After magnetization and a relaxation time of 900 s, we continue with cross-field demagnetization as follows:
\begin{itemize}
 \item We move the sample from the split coil electromagnet to the transverse applied field coil with maximum applied field ${B_{ax,max}\sim}$200 mT.
 \item We apply ripples of the magnetic field, ${B_{ax}}$, with different amplitudes, according to the trapped field ${B_t}$ as ${B_{ax}=B_t/2,B_t/4,B_t/8}$ and frequencies 0.1 and 1 Hz. 
Where ${B_t}$ is trapped field measured 100 ${\mu}$m above the top centre of the sample. 
 \item We measure the demagnetization for another 10 minutes.
\end{itemize}
The waveform of the applied magnetic fields is shown in Fig. \ref{Fields.fig}.

We used a lock-in amplifier to generate the AC signal and measured the voltage across the 0.5 mOhm resistor. The generated signal was amplified with two amplifiers to generate an AC current, which was 
passed through the coil to generate the AC field. The trapped field was measured by a Hall probe array of 7 sensors Multi-7U\cite{Arepoc} (Fig. \ref{Demag.fig}). The Hall-probe array covers only 3.5 mm 
of the sample, and hence measurements only provide partial information on the trapped field profile. The Hall probe array is at 100 ${\mu}$m above the sample surface.


\section{Modeling}

\begin{table}[!t]
\renewcommand{\arraystretch}{1.3}
\caption{Input parameters for calculation}
\label{input.tab}
\centering
\begin{tabular}{|c||c|}
\hline
Size[mm] & 6x6x6\\
\hline
${J_c}$[A/m${^2}$] & 2.6${\times 10^8}$\\
\hline
${B_{az,max}}$[T] & 1.3\\
\hline
Ramp rate[mT/s] & 13\\
\hline
Relaxation[s] & 900\\
\hline
${E_c}$[V/m] & 1e-4\\
\hline
${f_{ax}}$[Hz] & 0.1,1\\
\hline
${B_{ax}}$[mT] & 35,73,130\\
\hline
n[-] & 30\\
\hline
\end{tabular}
\end{table}

For both modelling methods, MEMEP and FEM, we use the parameters listed in Table \ref{input.tab}.
In the calculation we used the isotropic ${\vE(\vJ)}$ power law with n-value 30,  
\begin{equation}
{{\bf{E}}({\bf{J}})=E_{c}\left(\frac{|{\bf{J}}|}{J_{c}}\right)^{n}\frac{\bf {J}}{\bf {|J|}}}, \label{Power_Law} \\
\end{equation}  
being more realistic than the critical state model, and ${E_c}$ is the critical electric field and ${J_c}$ is the critical current density. The model assumes constant ${J_c}$. 

\subsection{MEMEP model}

The model is based on the Minimum Electro-Magnetic Entropy Production in 3D \cite{Pardo17JCP}. This is a variational method with ${\vT}$ formulation. We take the interpretation that the ${\vT}$ vector is the effective magnetization, and hence ${\vT}$ outside the sample is zero. Therefore, this method avoids discretization and calculation of variables outside the sample. For each time step, the minimum of that functional is unique. Moreover, MEMEP can also take anisotropic ${\vE(\vJ)}$ relations into account, such as those from force-free effects. The self-programmed modelling tool is written in C++ with BoostMPI commands for parallel computing on a computer cluster. Sector minimization \cite{Pardo17JCP} was used to both speed up and parallelize the calculations.      

\subsection{FEM model}

The finite element method is based on 3D $\textit{\textbf{H}}$-formulation \cite{zhangM12SSTa}, \cite{grilli13Cry}, \cite{Ainslie14SST}, \cite{Ainslie15SST} implemented in Comsol Multiphysics 5.2a. 
The $\textit{\textbf{H}}$-formulation is derived from Faraday's and Ampere's laws, and the nonlinear electrical resistivity of the superconductor is represented by the ${E(J)}$ power law. 
Isothermal conditions are assumed; hence, no thermal model is included.


\section{Results and discussion}

\subsection{Measurements}

\begin{figure}[tbp]
\centering
{\includegraphics[trim=0 34 10 0,clip,width=7 cm]{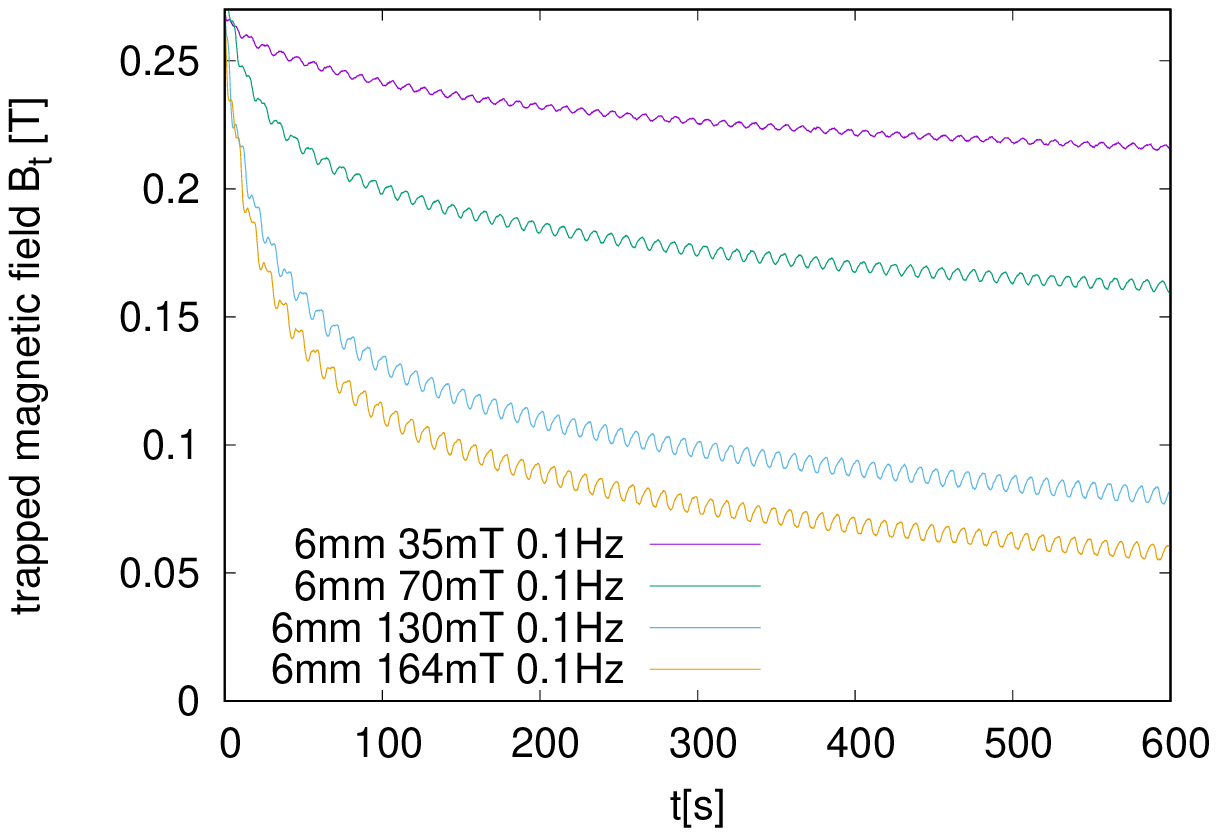}}

{\includegraphics[trim=0 0 10 10,clip,width=7 cm]{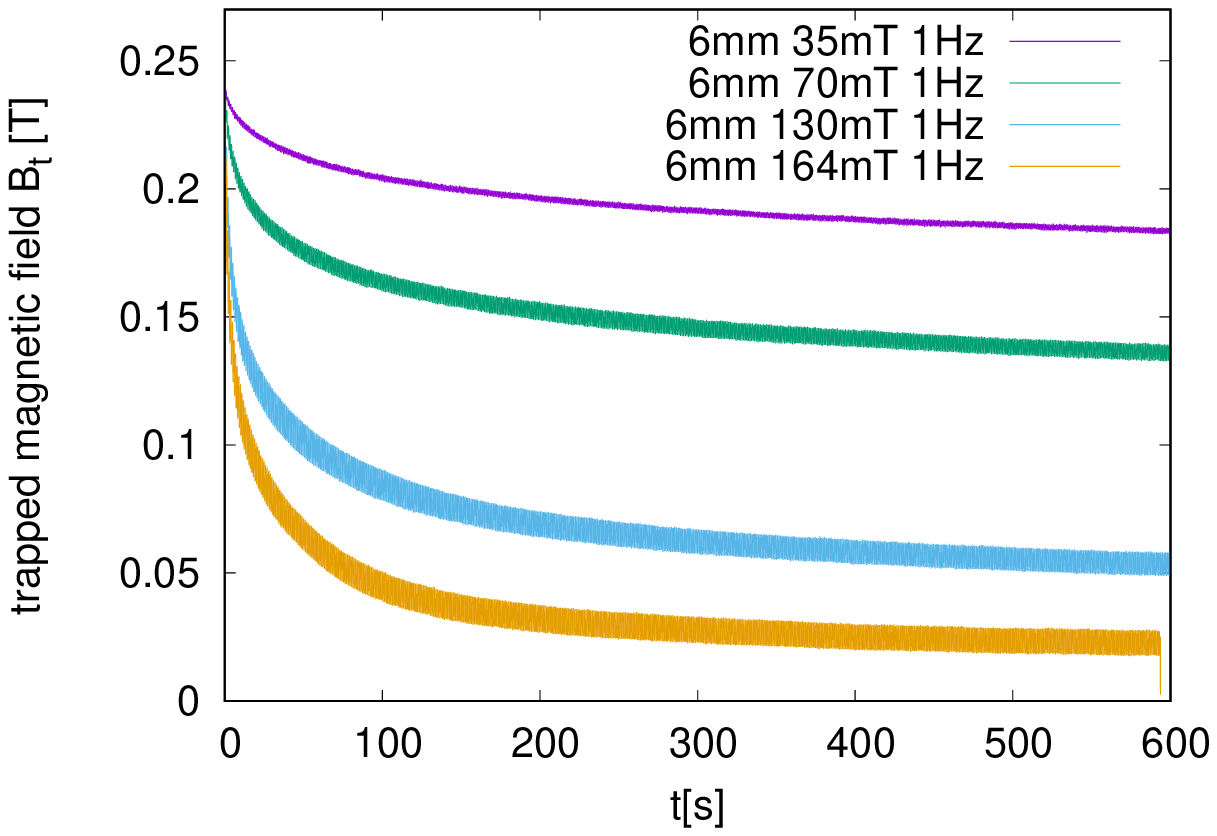}} 
\caption{Measured demagnetization of trapped field caused by a sinusoidal transverse applied field ${B_{ax}}$ 6 mm sample at 0.1 Hz(top), and 6 mm sample at 1 Hz(bottom).}
\label{Bt.fig}
\end{figure}

\begin{figure}[tbp]
\centering
{\includegraphics[trim=0 10 10 18,clip,width=7 cm]{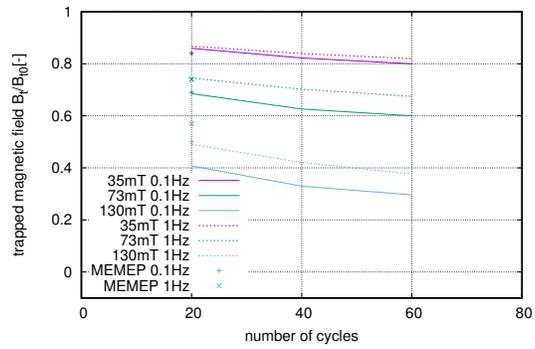}}
\caption{Measured dependence of trapped field on number of cycles of ripples for the 6 mm sample with calculation provided by MEMEP model. The figure is normalized by $B_{t0}$, which is the trapped field 
at the end of the relaxation time.}
\label{Cycles.fig}
\end{figure}

The 6 mm cubic sample was magnetized as described in Section II. The trapped field ${B_t}$ was 0.27 T. Demagnetization was done by ripples along the ${X}$ axis (Fig. \ref{Demag.fig}) 
with frequency 0.1 and 1 Hz and amplitude ${B_{ax}=B_t/2,B_t/4,B_t/8}$. The trapped field at the centre of the top surface of the cube is shown in Fig. \ref{Bt.fig}. During demagnetization, there appear 
ripples in the trapped field (Fig. \ref{Bt.fig}), which are slightly frequency dependent. The ripples increase with the transverse field amplitude. These ripples also appear in the models, although with lower amplitude.
The dependences of the trapped field on the number of demagnetizing cycles is on Fig. \ref{Cycles.fig}. The demagnetization is increasing with ripple amplitudes. There is a frequency 
dependence, of around 10${\%}$.
Applied fields of higher frequencies create higher induced electric fields in the sample, causing higher current densities and lower penetration depths \cite{sander10JCS,thakur11SSTa,thakur11SSTb,acreview}; and thence decreasing demagnetization per cycle.

\begin{figure}[tbp]
\centering
{\includegraphics[trim=0 0 10 0,clip,width=7 cm]{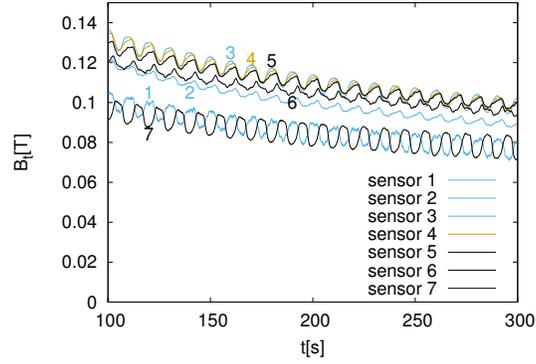}}
\caption{Hall-probe measurements of the 6 mm sample for ripple frequency of 0.1 Hz  ${B_t}$=0.27 T and ${B_{ax}}$=130 mT.}
\label{Sensors.fig}
\end{figure}

There is phase shift (180$\degree$) of trapped field for all measurements between the leftmost (1st) and rightmost (7th) Hall probe sensors. 
The most significant case is shown in Fig. \ref{Sensors.fig}. The oscillations and phase shift come from the applied field ripples $B_{ax}$. Both $J_y$ and $J_z$ caused by the applied ripples change their sign after each half-cycle (Fig. \ref{J.fig}), causing opposite contribution to the measured field.

\subsection{Modeling}

\begin{figure}[tbp]
\centering
{\includegraphics[trim=2 45 20 55,clip,width=8.8 cm]{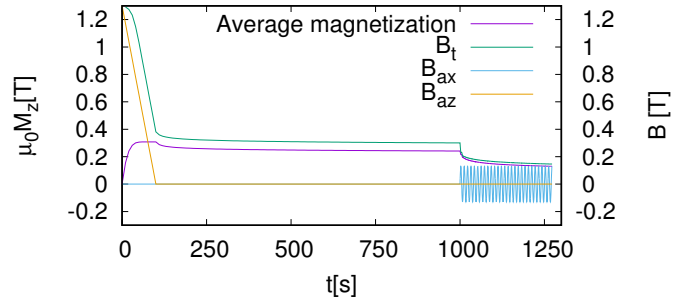}} 
\caption{Calculated magnetization of 6 mm wide sample. (a) The wave-form of applied fields, under triangular magnetizing field ${B_{az}}$ with peak 1.3 T and ramp rate 13 mT/s and 
sinusoidal ripples of frequency 0.1 Hz and ${B_{ax,{\rm max}}}$=130 mT.}
\label{Fields.fig}
\end{figure}

\begin{figure}[tbp]
\centering
{\includegraphics[trim=0 5 0 15,clip,width=7 cm]{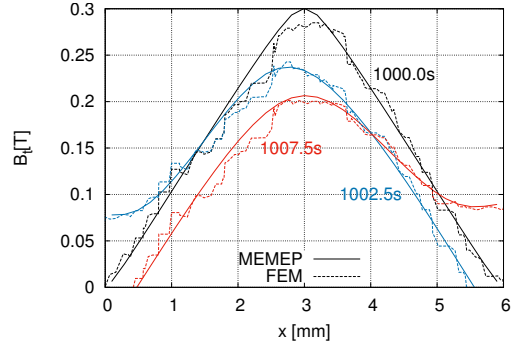}} 
\caption{Computed trapped field by both methods MEMEP (M) and FEM (F) after 15 minutes relaxation at 1000.0 s, at the positive peak (1002.5 s) of the ripple field 130 mT and at the negative peak (1007.5 s) 
of the ripple field -130mT.}
\label{Btall.fig}
\end{figure}

\begin{figure}[tbp]
\centering
{\includegraphics[trim=0 0 48 0,clip,height=3.0 cm]{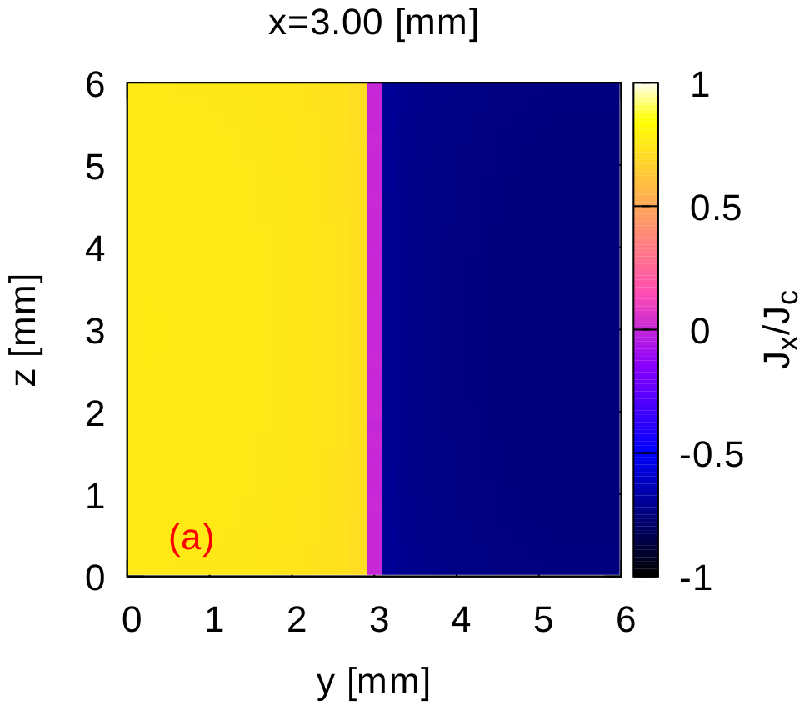}} 
{\includegraphics[trim=22 0 48 0,clip,height=3.0 cm]{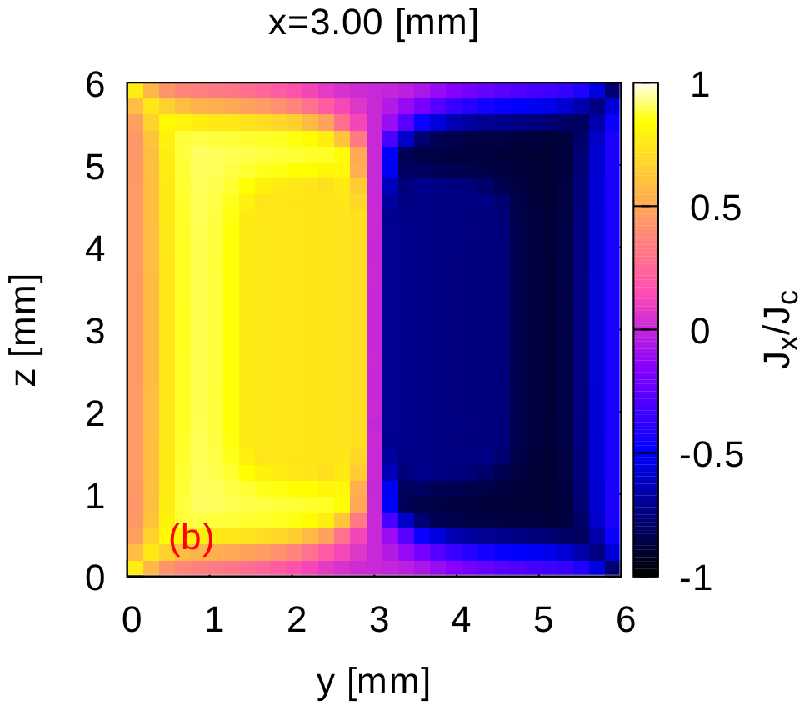}} 
{\includegraphics[trim=22 0 0 0,clip,height=3.0 cm]{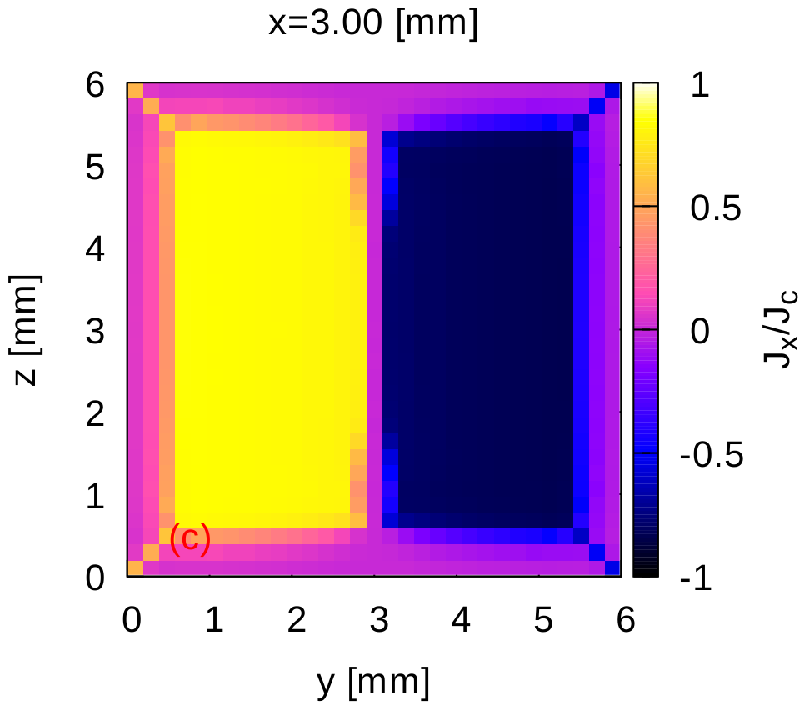}} \\
\vspace{1 mm}
{\includegraphics[trim=0 0 50 0,clip,height=3.0 cm]{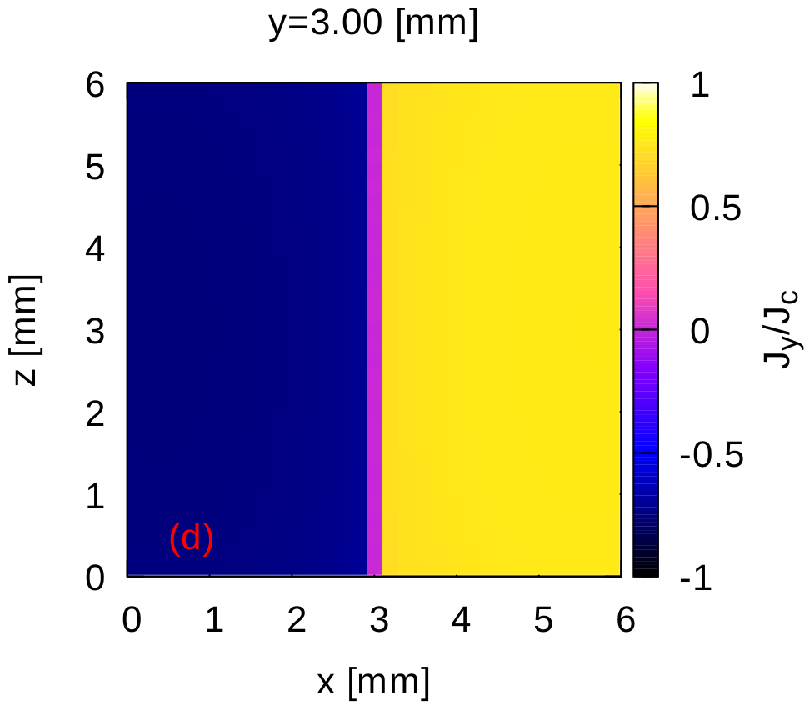}} 
{\includegraphics[trim=22 0 50 0,clip,height=3.0 cm]{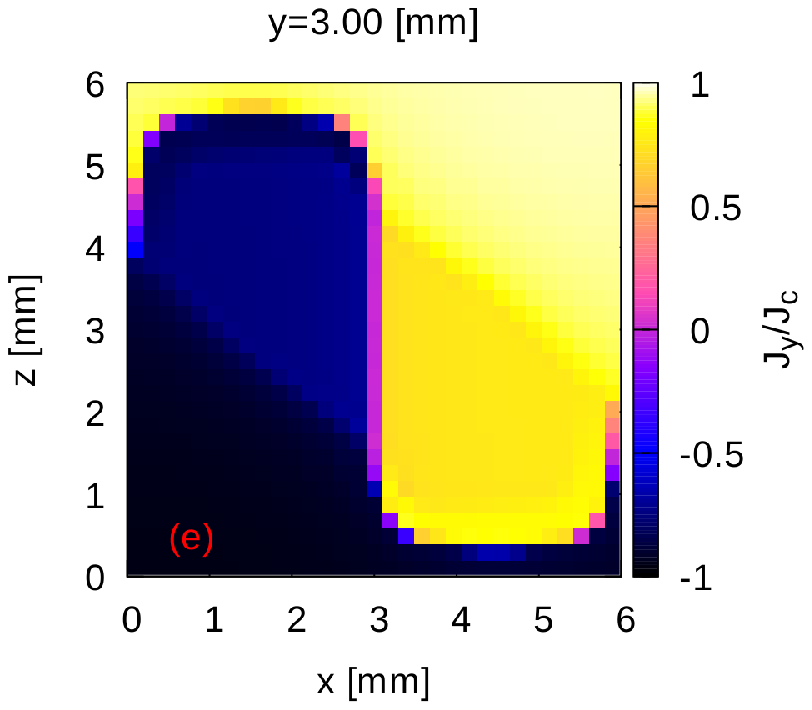}} 
{\includegraphics[trim=22 0 0 0,clip,height=3.0 cm]{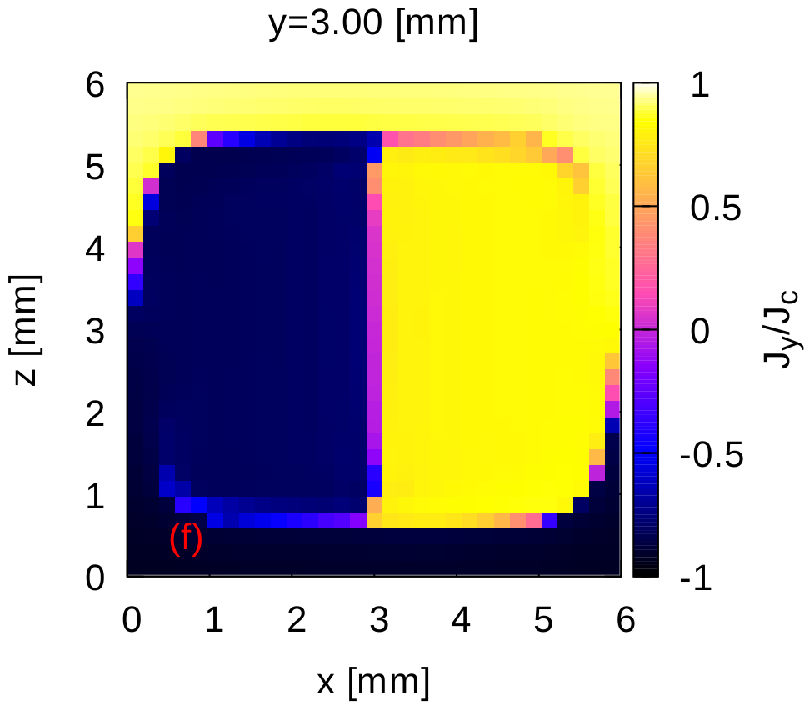}} \\
\vspace{1 mm}
{\includegraphics[trim=0 0 48 0,clip,height=3.0 cm]{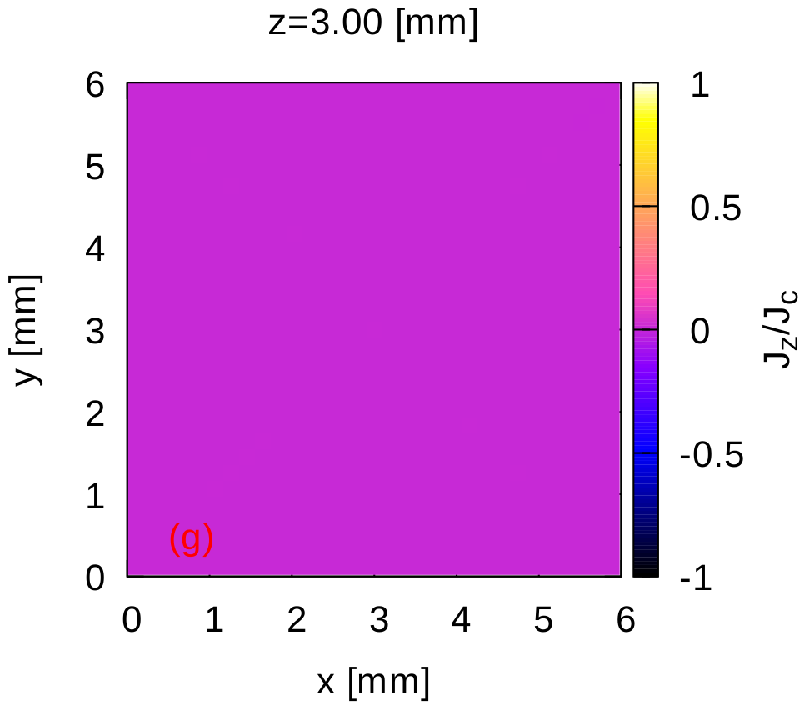}} 
{\includegraphics[trim=22 0 48 0,clip,height=3.0 cm]{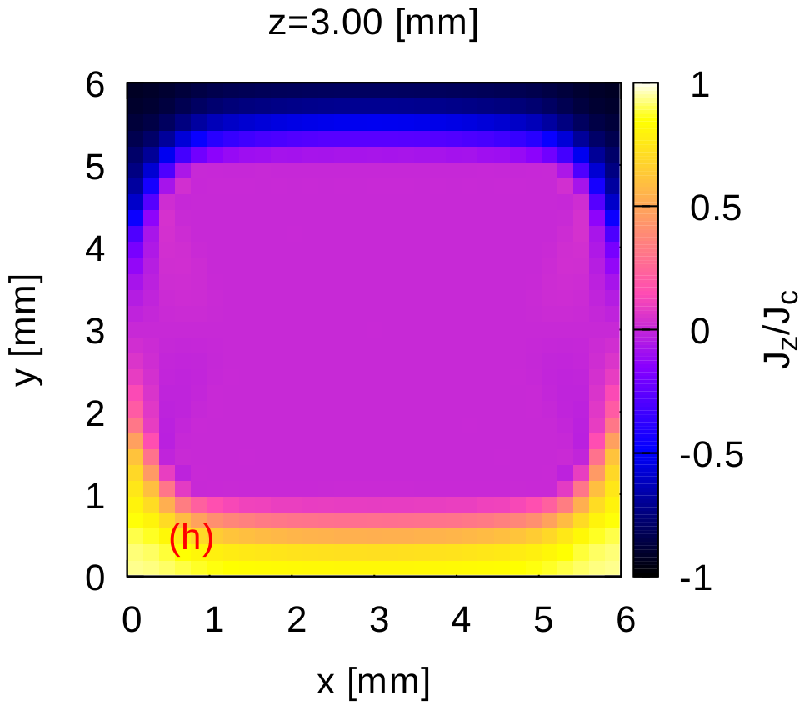}}
{\includegraphics[trim=22 0 0 0,clip,height=3.0 cm]{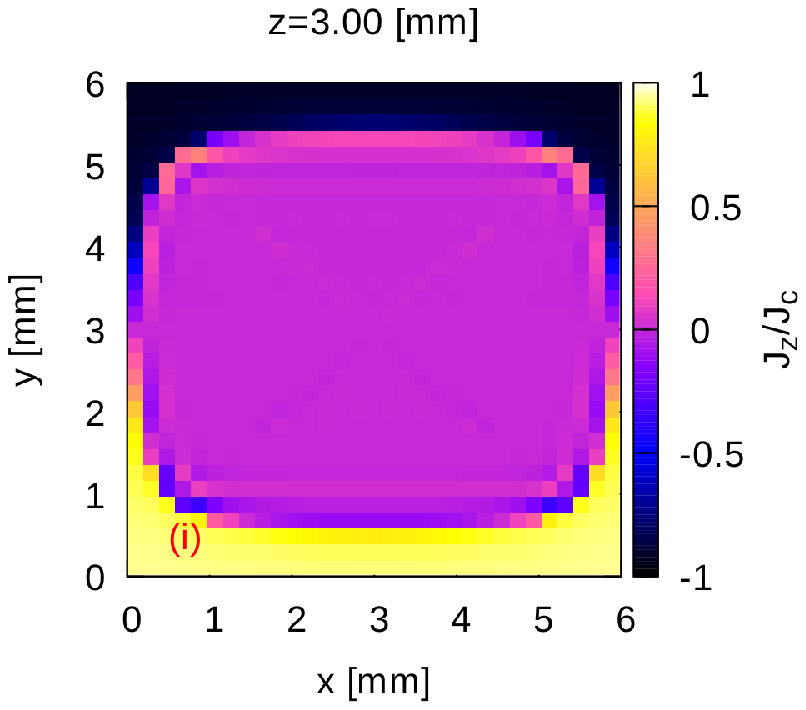}}
\caption{3D penetration of the current density to the cube, ${J_x,J_y,J_z}$ component of the current density cross-section at the middle of the cube x=3 mm, y=3 mm, z=3 mm (a,d,g) after 15 minutes relaxation 
(b,e,h) at first positive peak of ripple field 130 mT (c,f,i) at last positive peak	(10th cycle).}
\label{J.fig}
\end{figure}
\begin{figure}[tbp]
{\includegraphics[trim=40 0 50 10,clip,height=4.35 cm]{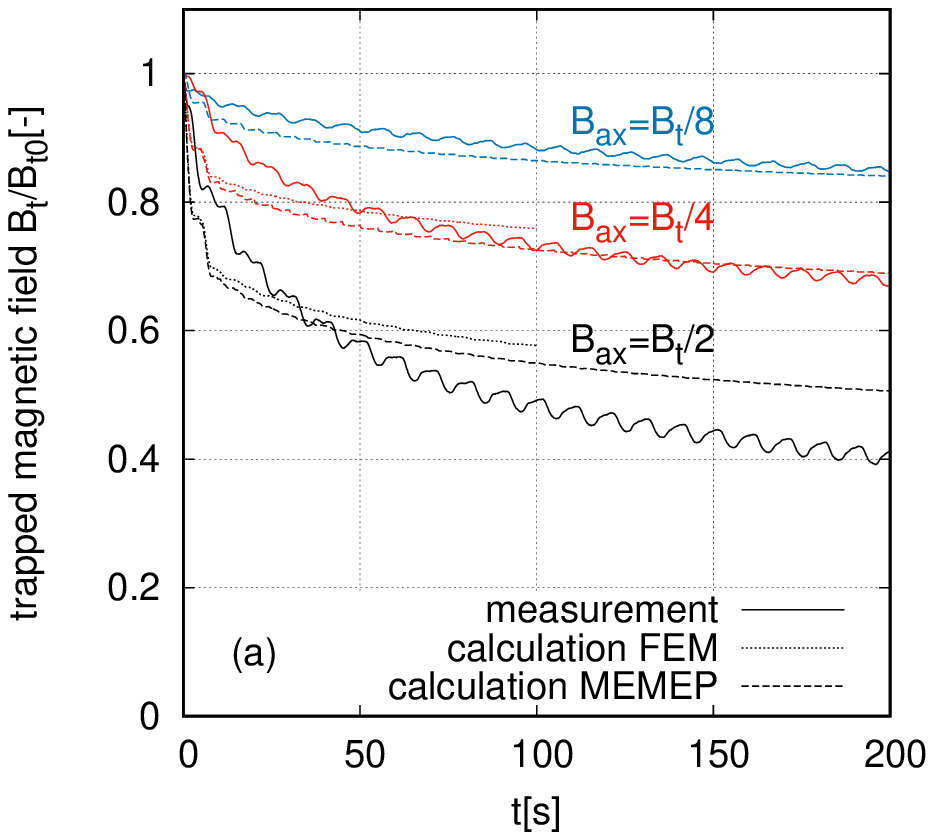}} 
{\includegraphics[trim=93 0 50 10,clip,height=4.35 cm]{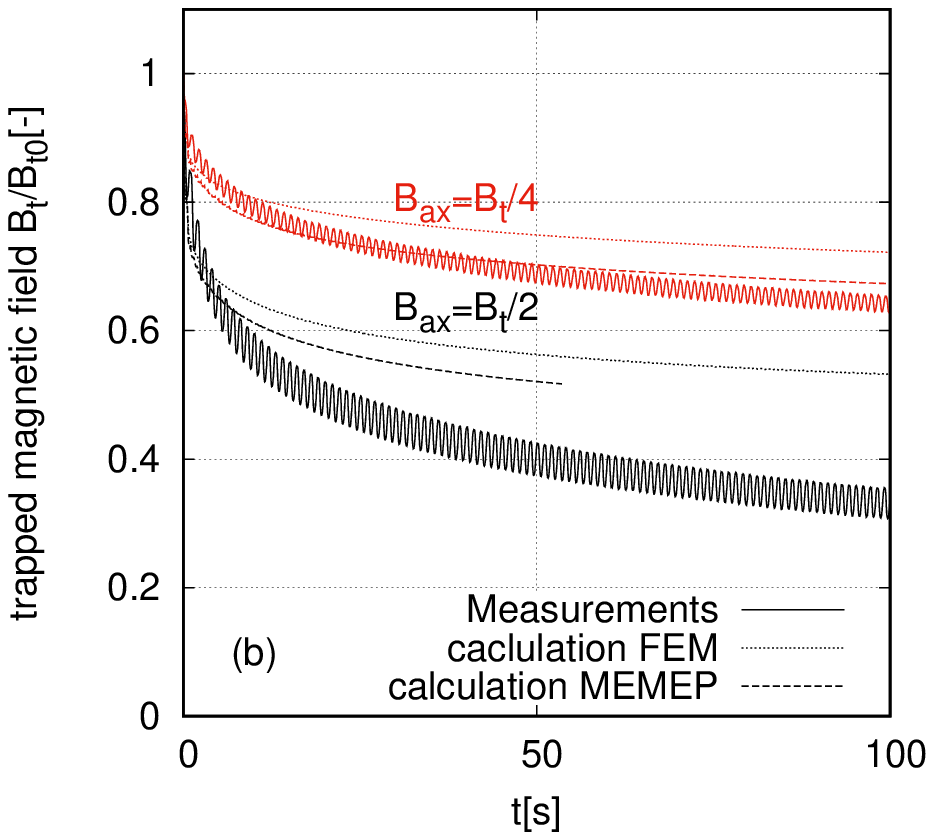}} 
\caption{Comparison of MEMEP, FEM and measurements with amplitude of ripples (a) ${B_{ax}/B_t=1/2,1/4,1/8}$ for 0.1 Hz and (b) ${B_{ax}/B_t=1/2,1/4}$ for 1 Hz. }
\label{Btcom.fig}
\end{figure}

\begin{figure}[tbp]
\centering
{\includegraphics[trim=0 0 0 20,clip,height=3.0 cm]{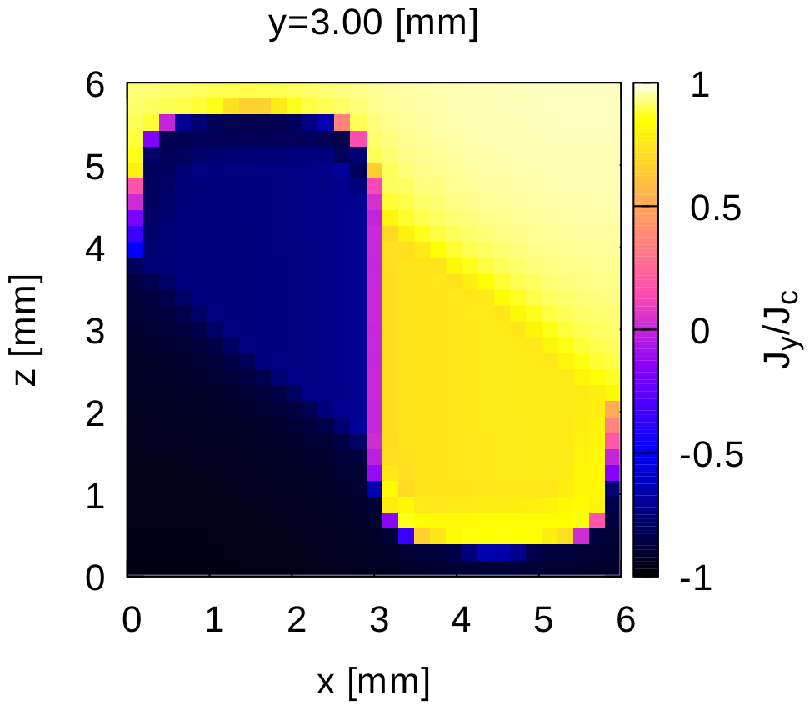}} 
{\includegraphics[trim=0 0 -10 0,clip,height=3 cm]{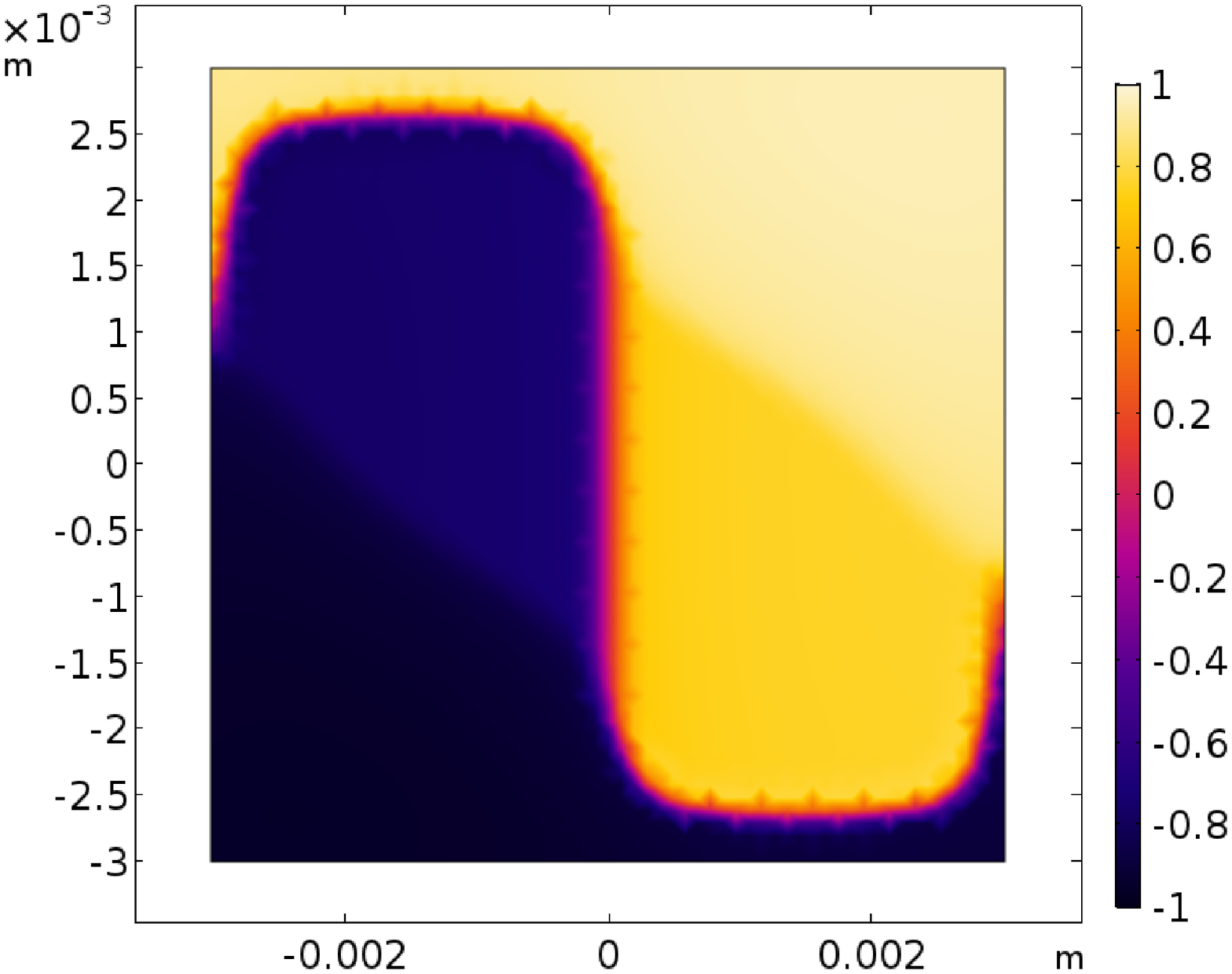}} 
\caption{Comparison of ${J_y}$ component at the mid plane at 1st positive peak at 1002.5 s of both methods: MEMEP (left) FEM (right).}
\label{Jy.fig}
\end{figure}

The calculated trapped field ${B_t}$ at the centre of the plane 100 ${\mu}$m above the sample and the average magnetization inside the sample during the entire demagnetization process is shown in Fig. 
\ref{Fields.fig} (and Fig. \ref{Btall.fig} and \ref{Btcom.fig}). We choose $J_c$ such that the calculated trapped field at the end of relaxation, of value ${B_t}$=0.3 T, corresponds to the measured one, being 
$J_c=2.6\times 10^8$ A/m$^2$.

The usual peak of the trapped field profile after relaxation is on Fig. \ref{Btall.fig}, curve at 1000 s. The sample is fully saturated, what confirmed the components of current density 
${J_x, J_y, J_z}$ at the mid-planes perpendicular to each component, Fig. \ref{J.fig}(a),(d),(g). As a result of full saturation, the ${J_z}$ component is almost zero [Fig. \ref{J.fig}(g)].
 
Next, we applied ripples parallel to the ${x}$ axis, which demagnetized the sample and caused a decrease in the trapped field (Fig. \ref{Fields.fig}). Already at the first cycle of the ripple, 
there appeared asymmetry of the trapped field (Fig. \ref{Btall.fig}) between the first positive and the first negative peak of ripple. Both methods showed good agreement. However, there are 
small inaccuracies due to the linear (first-order) elements and coarse mesh used in FEM model. 

The explanation of the asymmetry is in the shape of the current path. The screening current from the ripples changed the ${J_y}$ component into S-shape current fronts [Fig. \ref{J.fig}(e)], 
as it was explained in \cite{Campbell17SST} by 2D computations. ${J_x}$ close to the edges [Fig. \ref{J.fig}(b)] is erased by ${J_z}$ [Fig. \ref{J.fig}(h)] and both components present smooth current fronts. 
There is a difference of ${J_x, J_y}$ components between the first [Fig. \ref{J.fig}(b),(e)] and the last [Fig. \ref{J.fig}(c),(f)] cycle of ripples at the same 
instantaneous ${B_{ax}}$. The current density induced from ${B_{a,z}}$ decreases to a value below ${J_c}$ due to relaxation, which accounts for a small portion of the reduction of trapped field. 
The ${J_x}$ and ${J_z}$ components [Fig. \ref{J.fig}(c) and \ref{J.fig}(i)] present a sharp current front to zero. 

Next, we compare measurements, MEMEP and FEM (see Fig. \ref{Btcom.fig}). The models are with small deviance due to inaccuracies in the central trapped field after relaxation in FEM. 
For low amplitudes of ripples ${B_{ax}/B_t=1/8,1/4}$, we reach a nice agreement with measurements, but for higher fields like ${B_{ax}/B_t=1/2}$ the discrepancy increased. 
The models show steeper decrease of the trapped field during the first cycles, but for the following, calculations follow the measurements. Comparing demagnetization between 0.1 and 1 Hz for 
the same number of cycles (Fig. \ref{Cycles.fig}), calculations show the same qualitative behaviour as measurements, confirming that the cause of frequency dependence is the finite power-law exponent. 
We compared as well the ${J_y}$ component of current density profile between models (Fig. \ref{Jy.fig}), which agree very well and confirm the same behaviour of the models.

\section{Conclusion}

This article analyzed the demagnetization of GdBCO cubic bulks by transverse AC fields. The trapped field of a 6 mm sample was measured and calculated by two numerical methods, MEMEP 3D and FEM. 
The measurements confirm the asymmetry of the trapped field during the ripples with different amplitudes. The asymmetry comes from the 3D current paths inside the sample, which we explained by 3D model 
based on the MEMEP 3D variational method. The model showed reduction of ${|\vJ|}$ below ${J_c}$ in the places where ripples did not rewrite the previous state of current density. We saw as well sharp current fronts of current density after 10 cycles of ripples from positive to negative values of ${J}$. Both models show good consistency with each other. These models also agree with the measurements for low transverse fields, in spite of the simplification in the assumed superconductors properties (constant isotropic ${J_c }$ and n-value). Taking a more realistic n-value, the magnetic-field dependence of ${J_c}$ and anisotropy into account will provide better agreement also at high ripple transverse fields. The MEMEP 3D and FEM 3D models are useful tools to reveal all finite size effects of any model case and help explain 3D current paths.      

\section*{Acknowledgements}

M. Kapolka acknowledge the use of computing resources provided by the project SIVVP, ITMS 26230120002 supported by the Research \& Development Operational Programme funded by the ERDF, 
the financial support of the Grant Agency of the Ministry of Education of the Slovak Republic and the Slovak Academy of Sciences (VEGA) under contract No. 2/0126/15.

M. Ainslie would like to acknowledge financial support from a Royal Academy of Engineering Research Fellowship and an Engineering and Physical Sciences Research Council (EPSRC) Early Career Fellowship 
EP/P020313/1. All data are provided in full in the results section of this paper.



\begin{thebibliography}{10}
\providecommand{\url}[1]{#1}
\csname url@samestyle\endcsname
\providecommand{\newblock}{\relax}
\providecommand{\bibinfo}[2]{#2}
\providecommand{\BIBentrySTDinterwordspacing}{\spaceskip=0pt\relax}
\providecommand{\BIBentryALTinterwordstretchfactor}{4}
\providecommand{\BIBentryALTinterwordspacing}{\spaceskip=\fontdimen2\font plus
\BIBentryALTinterwordstretchfactor\fontdimen3\font minus
  \fontdimen4\font\relax}
\providecommand{\BIBforeignlanguage}[2]{{%
\expandafter\ifx\csname l@#1\endcsname\relax
\typeout{** WARNING: IEEEtran.bst: No hyphenation pattern has been}%
\typeout{** loaded for the language `#1'. Using the pattern for}%
\typeout{** the default language instead.}%
\else
\language=\csname l@#1\endcsname
\fi
#2}}
\providecommand{\BIBdecl}{\relax}
\BIBdecl

\bibitem{Durrell14SST}
J.~H. Durrell, A.~R. Dennis, J.~Jaroszynski, M.~D. Ainslie, K.~G.~B. Palmer,
  Y.~H. Shi, A.~C. Campbell, J.~Hull, M.~Strasik, E.~E. Hellstrom, and D.~A.
  Cardwell, ``A trapped field of 17.6t in melt-processed, bulk gd-ba-cu-o
  reinforced with shrink-fit steel,'' \emph{Supercond. Sci. Technol.}, vol.~27,
  p.~5, 2014.

\bibitem{Namburi16JECS}
D.~Namburi, Y.~Shi, K.~Palmer, A.~Dennis, J.~Durrell, and D.~Cardwell, ``An
  improved top seeded infiltration growth method for the fabrication of
  y–ba–cu–o bulk superconductors,'' \emph{J. Euro. Cer. Soc.}, vol.~36,
  pp. 615--624, 2016.

\bibitem{Namburi16SST}
------, ``A novel, two-step top seeded infiltration and growth process for the
  fabrication of single grain, bulk (re)bco superconductors,'' \emph{Supercond.
  Sci. Technol.}, vol.~29, p.~11, 2016.

\bibitem{Escamez16IES}
G.~Escamez, F.~Sirois, V.~Lahtinen, A.~Stenvall, A.~Badel, P.~Tixador,
  B.~Ramdane, G.~Meunier, R.~Perrin-Bit, and C.~E. Bruzek, ``3-d numerical
  modeling of ac losses in multifilamentary mgb 2 wires,'' \emph{IEEE Trans.
  Appl. Supercond.}, vol.~26, no.~3, pp. 1--7, 2016.

\bibitem{zermeno14SSTa}
V.~M.~R. Zermeno and F.~Grilli, ``{3D} modeling and simulation of {2G HTS}
  stacks and coils,'' \emph{Supercond. Sci. Technol.}, vol.~27, p. 044025,
  2014.

\bibitem{grilli13Cry}
F.~Grilli, R.~Brambilla, F.~Sirois, A.~Stenvall, and S.~Memiaghe, ``Development
  of a three-dimensional finite-element model for high-temperature
  superconductors based on the {$H$}-formulation,'' \emph{Cryogenics}, vol.~53,
  pp. 142--147, 2013.

\bibitem{Farinon14SST}
S.~Farinon, G.~Iannone, P.~Fabbricatore, and U.~Gambardella, ``{2D} and {3D}
  numerical modeling of experimental magnetization cycles in disks and
  spheres,'' \emph{Supercond. Sci. Technol.}, vol.~27, no.~10, p. 104005, 2014.

\bibitem{Grilli05IES}
F.~Grilli, S.~Stavrev, Y.~Le~Floch, M.~Costa-Bouzo, E.~Vinot, I.~Klutsch,
  G.~Meunier, P.~Tixador, and B.~Dutoit, ``Finite-element method modeling of
  superconductors: from {2-D} to {3-D},'' \emph{IEEE Trans. Appl. Supercond.},
  vol.~15, no.~1, pp. 17--25, 2005.

\bibitem{Stenvall14SST}
A.~Stenvall, V.~Lahtinen, and M.~Lyly, ``An {H-formulation-based}
  three-dimensional hysteresis loss modelling tool in a simulation including
  time varying applied field and transport current: the fundamental problem and
  its solution,'' \emph{Supercond. Sci. Technol.}, vol.~27, no.~10, p. 104004,
  2014.

\bibitem{Bossavit94IEG}
A.~Bossavit, ``Numerical modelling of superconductors in three dimensions: a
  model and a finite element method.''

\bibitem{badia01PRL}
A.~Bad{\'\i}a and C.~L{\'o}pez, ``Critical state theory for nonparallel flux
  line lattices in {type-II} superconductors,'' \emph{Phys. Rev. Lett.},
  vol.~87, no.~12, p. 127004, 2001.

\bibitem{Elliot06JMA}
C.~Elliott and Y.~Kashima, ``A finite-element analysis of critical-state models
  for type-ii superconductivity in 3d,'' \emph{J. Numer. Anal.}, vol.~27, pp.
  293--331, 2006.

\bibitem{Pardo17JCP}
E.~Pardo and M.~Kapolka, ``3d computation of non-linear eddy currents:
  variational method and superconducting cubic bulk,'' \emph{J. Comput. Phys.},
  2017.

\bibitem{prigozhin97IES}
L.~Prigozhin, ``Analysis of critical-state problems in type{-II}
  superconductivity,'' \emph{IEEE Trans. Appl. Supercond.}, vol.~7, no.~4, pp.
  3866--3873, 1997.

\bibitem{Pardo15SST}
E.~Pardo, J.~{\v Souc}, and L.~{Frolek}, ``Electromagnetic modelling of
  superconductors with a smooth current-voltage relation: variational principle
  and coils from a few turns to large magnets,'' \emph{Supercond. Sci.
  Technol.}, vol.~28, p. 044003, 2015.

\bibitem{Vanderbemden03IES}
P.~Vanderbemden, S.~Dorbolo, N.~Hari-Babu, A.~Ntatsis, and D.~C.~A. Campbell,
  ``Behavior of bulk melt-textured ybco single domains subjected to crossed
  magnetic fields,'' \emph{IEEE Trans. Appl. Supercond.}, vol.~13, pp.
  3746--3749, 2003.

\bibitem{vanderbemden07SST}
P.~Vanderbemden, Z.~Hong, T.~Coombs, M.~Ausloos, N.~Hari~Babu, D.~Cardwell, and
  A.~Campbell, ``Remagnetization of bulk high-temperature superconductors
  subjected to crossed and rotating magnetic fields,'' \emph{Supercond. Sci.
  Technol.}, vol.~20, p. S174, 2007.

\bibitem{Fagnard16SST}
J.~F. Fagnard, M.~Morita, S.~Nariki, H.~Teshima, H.~Caps, B.~Vanderheyden, and
  P.~Vanderbemden, ``Magnetic moment and local magnetic induction of
  superconducting/ferromagnetic structures subjected to crossed fields:
  experiments on gdbco and modeling,'' \emph{Supercond. Sci. Technol.},
  vol.~29, 2016.

\bibitem{Campbell17SST}
A.~Campbell, M.~Baghdadi, A.~Pafel, D.~Zhou, K.~Y. Huang, Y.~Shi, and
  T.~Coombs, ``Demagnetisation by crossed fields in superconductors,''
  \emph{Supercond. Sci. Technol.}, vol.~30, 2017.

\bibitem{Ainslie15SST}
M.~D. Ainslie and F.~Fujishiro, ``Modelling of bulk superconductor
  magnetization,'' \emph{Supercond. Sci. Technol.}, vol.~28, 2015.

\bibitem{Arepoc}
Multi-7U,available online at: {http://www.arepoc.sk/?p=29}.

\bibitem{zhangM12SSTa}
M.~Zhang and T.~Coombs, ``{3D} modeling of high{-$T_c$} superconductors by
  finite element software,'' \emph{Supercond. Sci. Technol.}, vol.~25, p.
  015009, 2012.

\bibitem{Ainslie14SST}
M.~D. Ainslie, H.~Fujishiro, T.~Ujiie, J.~Zou, A.~R. Dennis, Y.-H. Shi, and
  D.~A. Cardwell, ``Modelling and comparison of trapped fields in (re)bco bulk
  superconductors for activation using pulsed field magnetization,''
  \emph{Supercond. Sci. Technol.}, vol.~27, no. 065008, p.~9, 2014.

\bibitem{sander10JCS}
M.~Sander and F.~Grilli, ``{FEM-}calculations on the frequency dependence of
  hysteretic losses in coated conductors,'' vol. 234, no.~2, p. 022030, 2010.

\bibitem{thakur11SSTa}
K.~P. Thakur, A.~Raj, E.~H. Brandt, and P.~V. Sastry, ``Frequency dependent
  magnetization of superconductor strip,'' \emph{Supercond. Sci. Technol.},
  vol.~24, no.~4, p. 045006, 2011.

\bibitem{thakur11SSTb}
K.~P. Thakur, A.~Raj, E.~H. Brandt, J.~Kvitkovic, and S.~V. Pamidi,
  ``Frequency-dependent critical current and transport ac loss of
  superconductor strip and roebel cable,'' \emph{Supercond. Sci. Technol.},
  vol.~24, no.~6, p. 065024, 2011.

\bibitem{acreview}
F.~Grilli, E.~Pardo, A.~Stenvall, D.~N. Nguyen, W.~Yuan, and F.~G{\"om\"o}ry,
  ``Computation of losses in {HTS} under the action of varying magnetic fields
  and currents,'' \emph{IEEE Trans. Appl. Supercond.}, vol.~24, no.~1, p.
  8200433, 2014.

\end{thebibliography}

\end{document}